\documentclass[a4paper,11pt]{article}
\usepackage{amssymb}

\def\d{\partial}

\newcommand{\beq}{\begin{equation}}
\newcommand{\eeq}{\end{equation}}
\newcommand{\bg}{\begin{gather}}
\newcommand{\eg}{\end{gather}}
\newcommand{\CC}{\mathbb C}
\newcommand{\RR}{\mathbb R}
\newcommand{\ZZ}{\mathbb Z}
\newcommand{\NN}{\mathbb N}

\newcommand{\supp}{{\rm supp\>}}

\begin{document}
\title{Perturbative renormalization and BRST}
\author{Michael D\"utsch\\[2mm]
Institut f\"ur Theoretische Physik\\
Universit\"at Z\"urich\\
CH-8057 Z\"urich, Switzerland\\
{\tt \small duetsch@physik.unizh.ch}\\[2mm]
\and Klaus Fredenhagen \\[2mm]
II. Institut f\"ur Theoretische Physik\\
Universit\"at Hamburg\\
D-22761 Hamburg, Germany\\
{\tt \small klaus.fredenhagen@desy.de}}

\date{}

\maketitle
\section{Main problems in the perturbative quantization of gauge theories}
\setcounter{equation}{0}
Gauge theories are field theories in which the basic fields are not 
directly observable. Field configurations yielding the same observables 
are connected by a gauge transformation. In the classical theory
the Cauchy problem is well posed for the observables, but in general 
not for the nonobservable gauge variant basic fields, due to the existence 
of time dependent gauge transformations.

Attempts to quantize the gauge invariant objects directly have not yet 
been completely satisfactory. Instead one modifies the
classical action by adding a {\it gauge fixing term}
such that standard techniques of perturbative quantization can 
be applied and such that 
the dynamics of the gauge invariant classical fields is not changed. 
In perturbation theory this problem shows up already in the 
quantization of the free gauge fields (Sect.~3). In 
the final (interacting) theory the physical quantities should be 
independent on how the gauge fixing is done
('gauge independence').

Traditionally, the quantization 
of gauge theories is mostly analyzed in 
terms of path integrals (e.g.~by Faddeev and Popov) where some 
parts of the arguments are only heuristic. In 
the original treatment of Becchi, Rouet and Stora (cf.~also Tyutin)
(which is called 'BRST-quantization'),
a restriction to purely massive theories was necessary; 
the generalization to the massless 
case by Lowenstein's method is cumbersome.

The BRST-quantization is based on earlier work of 
Feynman, Faddeev and Popov (introduction of 
``ghost fields''), and of Slavnov.
The basic idea is that after adding a term to the 
Lagrangian which makes the Cauchy 
problem well posed but which is not gauge invariant 
one enlarges the number of fields by 
infinitesimal gauge tranformations (``ghosts'') and their 
duals (``anti-ghosts''). One then 
adds a further term to the Lagrangian which contains a coupling of the 
anti-ghosts and ghosts. 
The BRST transformation acts as an infinitesimal 
gauge transformation on the original fields and on the gauge transformations 
themselves
and maps the anti-ghosts to the gauge fixing terms.
This is done in such a way that the total 
Lagrangian is invariant and that the BRST transformation is nil-potent. 
The hard problem in the perturbative construction of gauge theories
is to show that BRST-symmetry can be maintained during renormalization 
(Sect.~4). By
means of the 'Quantum Action Principle' of Lowenstein (1971) and 
Lam (1972-1973) a cohomological classification of anomalies
was worked out (an overview is e.g.~given in the 
book of Piguet and Sorella (1995)).
For more details see $\to$ 'BRST Quantization'.  

The BRST-quantization can be carried out in a transparent way in 
the framework of 
Algebraic Quantum Field Theory ('AQFT', see $\to$
'Algebraic approach to quantum field theory'). 
The advantage of this formulation is that it allows 
to separate the three main problems of perturbative gauge theories:

\begin{itemize}

\item the elimination of unphysical degrees of freedom,

\item positivity (or ``unitarity'') 

\item and the problem of infrared divergences.

\end{itemize}

In AQFT, the procedure is the following: starting from an 
algebra of all local fields, 
including the unphysical ones, one shows that after 
perturbative quantization the algebra 
admits the BRST transformation as a graded nil-potent derivation. 
The algebra of observables 
is then defined as the cohomology of the BRST transformation. 
To solve the problem of positivity, one has to show that the algebra of 
observables, in contrast to the algebra of all fields, has 
a nontrivial representation on a Hilbert space. Finally, one can attack the 
infrared problem 
by investigating the asymptotic behavior of states. The latter 
problem is nontrivial 
even in quantum electrodynamics (since an electron is accompanied 
by a 'cloud of soft photons')
and may be related to confinement in 
quantum chromodynamics.   

The method of BRST quantization is by no means restricted to gauge 
theories, but applies to general constrained systems. In particular,
massive vector fields, where the masses are usually generated by the 
Higgs mechanism, can alternatively be treated directly by the BRST 
formalism, in close analogy to the massless case, cf. Sect.~3.

\section{Local operator BRST-formalism}\setcounter{equation}{0}
In AQFT, the principal object is the family of operator algebras
${\cal O}\rightarrow {\cal A}({\cal O})$ (where ${\cal O}$ runs e.g.~through all double 
cones in Minkowski space), which fulfills the Haag-Kastler axioms (cf.
$\to$ 'Algebraic approach to quantum field theory'). To construct 
these algebras one considers the algebras ${\cal F}({\cal O})$ which 
are generated by all 
local fields including ghosts $u$ and anti-ghosts $\tilde{u}$. Ghosts and anti-ghosts are scalar fermionic fields. The algebra gets a $\ZZ_2$ grading with respect to even and odd ghost numbers, where ghosts get ghost numbers $+1$ and 
anti-ghosts ghost number $-1$. 
The BRST-transformation $s$ acts on these algebras as  
a $\ZZ_2$-graded derivation
with $s^2=0$, $s({\cal F}({\cal O}))\subset {\cal F}({\cal O})$
and $s(F^*)=-(-1)^{\delta_F}s(F)^*$, $\delta_F$ denoting the ghost 
number of $F$.

The observables should be $s$-invariant and may be identified if they differ 
by a field in the range of $s$. 
Since the range ${\cal A}_{00}$ of $s$ is an ideal in the 
kernel ${\cal A}_0$ of $s$, 
the algebra of observables
is defined as the quotient
\begin{equation}
{\cal A}:= {\cal A}_0 / {\cal A}_{00} \ ,
\label{A}
\end{equation}
and the local algebras ${\cal A}({\cal O})\subset {\cal A}$ are the 
images of ${\cal A}_0\cap {\cal F}({\cal O})$ under the quotient map 
${\cal A}_0\rightarrow {\cal A}$.

To prove that ${\cal A}$ admits a nontrivial representation by 
operators on a Hilbert space one may use the BRST-operator formalism 
(Kugo - Ojima (1979) and D\"utsch - Fredenhagen (1999)):
one starts from  a representation of ${\cal F}$  on 
an inner product space $({\cal K},\langle\cdot ,\cdot\rangle)$ such that
$\langle F^*\phi,\psi\rangle=\langle\phi,F\psi\rangle$ and that $s$ is 
implemented by an operator 
$Q$ on ${\cal K}$, i.e.
\begin{equation}
s(F)=[Q,F],\label{[Q,F]}
\end{equation}
with $[\cdot,\cdot]$ denoting the graded commutator,
such that $Q$ is symmetric and nil-potent. 
One may then construct the space of physical states as the cohomology of $Q$, 
${\cal H}:= {\cal K}_0 / {\cal K}_{00}$ where ${\cal K}_0$ is the kernel and 
${\cal K}_{00}$ the range of $Q$.
The algebra of observables now has a natural representation $\pi$ on ${\cal H}$:
\begin{equation}
\pi ([A])[\phi]:= [A\phi]\label{pi}
\end{equation}
(where $A\in {\cal A}_0,\,\phi\in {\cal K}_0,\,[A]:=A+{\cal A}_{00},
\,[\phi]:=\phi+{\cal K}_{00}$). The crucial 
question is whether the scalar product on ${\cal H}$ inherited from 
${\cal K}$ is positive definite. 

In free quantum field theories $({\cal K},\langle\cdot ,\cdot\rangle)$
can be chosen in such a way that 
the positivity can directly be checked by identifying 
the physical degrees of freedom (Sect.~3). 
In interacting theories (Sect.~5) one may argue in terms of 
scattering states that the free BRST operator on the asymptotic fields coincides with the
BRST operator of the interacting theory. This argument, however, is invalidated by 
infrared problems in massless gauge theories. Instead one may use a
stability property of the construction.

Namely, let $\tilde{\cal F}$ be the algebra of formal power series with values in ${\cal F}$, 
and let $\tilde{\cal K}$ be the vector space of formal power series with 
values in ${\cal K}$. $\tilde{\cal K}$ possesses a natural inner product with values in 
the ring of formal power series $\CC[[\lambda]]$, as well as a representation of 
$\tilde{\cal F}$ by operators. One also assumes that the BRST transformation $\tilde{s}$ is 
a formal power series $\tilde{s}=\sum_n \lambda^n s_n$ of operators $s_n$ on ${\cal F}$
and that the BRST operator $\tilde{Q}$ is a formal power series 
$\tilde{Q}=\sum_n \lambda^n Q_n$ of operators on ${\cal K}$. The algebraic construction 
can then be done in the same way as before yielding a representation $\tilde{\pi}$ 
of the algebra of 
observables $\tilde{\cal A}$ by endomorphisms of a $\CC[[\lambda]]$ module 
$\tilde{\cal H}$, which has an 
inner product with values in $\CC[[\lambda]]$.

One now assumes that at $\lambda=0$ the inner product is positive, in the sense that 
\begin{eqnarray}
{\bf (Positivity)}\quad 
&{\rm (i)}&\quad \langle\phi,\phi \rangle \geq 0\quad\quad\forall 
\phi\in {\cal K}\  {\rm  with } \  Q_0\phi=0\ ,\nonumber\\
{\rm and}\quad &{\rm (ii)}&\quad Q_0\phi=0 \quad 
\wedge\quad \langle\phi,\phi\rangle =0\quad\Longrightarrow
\quad \phi\in {Q_0\cal K}\ .\label{positivity}
\end{eqnarray}

Then the inner product on $\tilde{\cal H}$ is positive in the sense that for all 
$\tilde{\phi}\in \tilde{\cal H}$ the inner product with itself,
$\langle \tilde{\phi},\tilde{\phi}\rangle$, is of the form $\tilde{c}^*\tilde{c}$ with some 
power series $\tilde{c}\in \CC[[\lambda]]$, and $\tilde c=0$ iff $\tilde\phi =0$. 

This result guarantees that, within perturbation theory, the interacting 
theory satisfies positivity, provided the unperturbed theory was positive 
and BRST symmetry is preserved.  
\section{Quantization of free gauge fields}\setcounter{equation}{0}
The action of a {\it classical} free gauge field $A$,
\beq
S_0(A)=-\frac{1}{4}\int dx\, F^{\mu\nu}(x)F_{\mu\nu}(x)=
\frac{1}{2}\int dk\,\hat{A}_\mu(k)^*M^{\mu\nu}(k)\hat{A}_\nu(k)\label{S}
\eeq    
(where $F^{\mu\nu}:=\d^\mu A^\nu-\d^\nu A^\mu$ and $M^{\mu\nu}(k):=k^2g^{\mu\nu}
-k^\mu k^\nu$) is unsuited for quantization because $M^{\mu\nu}$ is not invertible:
due to $M^{\mu\nu}k_\mu =0$ it has an eigenvalue $0$. Therefore, the action is usually 
modified by adding a Lorentz invariant {\it gauge fixing term}: $M^{\mu\nu}$ is replaced by
$M^{\mu\nu}(k)+\lambda k^\mu k^\nu$ where $\lambda\in\RR\setminus\{0\}$ is an 
arbitrary constant. The corresponding Euler-Lagrange equation reads
\beq
\square A^\mu - (1-\lambda)\d^\mu\d_\nu A^\nu=0\ .\label{ffeq}
\eeq
For simplicity let us choose $\lambda =1$, which is referred to as Feynman gauge. Then 
the algebra of the free gauge field is the unital $\star$-algebra generated by elements 
$A^\mu(f),\> f\in{\cal D}(\RR^4)$, which fulfill the relations:
\begin{eqnarray}
&&f\mapsto A^\mu(f)\quad\mathrm{ is\> linear}\ ,\\
&&A^\mu(\square f)=0\ ,\label{wave-eq}\\
&&A^\mu(f)^{*}=A^\mu(\bar f)\ ,\label{A^*}\\
&&[A^\mu(f),A^\nu(g)]=ig^{\mu\nu}\int dx\,dy\,f(x)D(x-y)g(y)\label{[A,A]}
\end{eqnarray}
where $D$ is the massless Pauli-Jordan distribution.

This algebra does not possess Hilbert space representations 
which satisfy the microlocal spectrum condition, a condition 
which in particular requires the singularity of the 2-point function 
to be of the so-called Hadamard form. It possesses instead representations
on vector spaces with a nondegenerate sequilinear form, e.g. the Fock 
space over a one particle space with scalar product
\beq
\langle \phi,\psi \rangle = (2\pi)^{-3}
\int \frac{d^3\vec{p}}{2|\vec{p}|}\, \overline{\phi^\mu(p)}\,
\psi_\mu(p)\vert_{p^0=|\vec{p}|}\ .
\eeq
Gupta and Bleuler characterized a subspace of the Fock space on 
which the scalar product is semidefinite; the space of physical states 
is then obtained by dividing out the space of vectors with vanishing norm.

After adding a mass term $\frac{m^2}{2}\int dx\, A_\mu(x)A^\mu(x)$  
to the action (\ref{S}), it seems to be no longer necessary to add also
a gauge fixing term. The fields then satisfy the Proca equation 
\beq
\d_\mu F^{\mu\nu}+m^2 A^\nu =0\ ,
\eeq
which is equivalent to $(\square+m^2) A^\mu=0$ together with the constraint 
$\d_\mu A^\mu =0$. The 
Cauchy problem is well posed, and the fields
can be represented in a positive 
norm Fock space with only physical states (corresponding to the three physical 
polarizations of $A$).
The problem, however, is that the corresponding propagator admits no 
power counting renormalizable perturbation series. 

The latter problem can be circumvented in the following way:
For the algebra of the free quantum field one takes only
$(\square+m^2) A^\mu=0$ into account (or equivalently
one adds the 'gauge fixing term' $\frac{1}{2}(\d_\mu A^\mu)^2$
to the Lagrangian) and
goes over from the physical field $A^\mu$ to
\beq
B^\mu :=A^\mu+\frac{\d^\mu\phi}{m}\ ,
\eeq
where $\phi$ is a real scalar field to the same mass $m$ where the sign 
of the commutator is reversed
('bosonic ghost field' or 'St\"uckelberg field'). The propagator of $B^\mu$
yields a power counting renormalizable perturbation series, however
$B^\mu$ is an unphysical field. One  
obtains four independent 
components of $B$
which satisfy the Klein Gordon equation. The constraint 
$0=\d_\mu A^\mu =\d_\mu B^\mu +m\phi$ is required for 
the expectation values in physical states only.
So quantization in the case $m>0$ can be treated in analogy to
(\ref{wave-eq})-(\ref{[A,A]}) by replacing $A^\mu$ by $B^\mu$,
the wave operator
by the Klein Gordon operator $(\square +m^2)$ in (\ref{wave-eq}) and 
$D$ by the 
corresponding massive commutator distribution $\Delta_m$ in (\ref{[A,A]}).
Again the algebra can be nontrivially represented on a space 
with indefinite metric, but not on a Hilbert space.

One can now use the method of  BRST quantization in the massless as 
well as in the massive case. One introduces a pair of fermionic 
scalar fields ('ghost fields') $(u,\tilde u)$. 
$u,\>\tilde u$ and (for $m>0$) $\phi$ fulfil the Klein Gordon equation
to the same mass $m\geq 0$ as the vector field $B$.
The free BRST-transformation reads 
\beq
s_0(B^\mu)=i\d^\mu u\ ,\quad s_0(\phi)=im u\ ,
\quad s_0(u)=0\,\quad s_0(\tilde u)=-i(\d_\nu B^\nu +m\phi) ,\label{free-BRS}
\eeq
see e.g.~the second book of G.~Scharf in the list below.
It is implemented by the free BRST-charge 
\beq
Q_0=\int_{x^0={\rm const.}} d^3x\,  j_0^{(0)}(x^0,\vec{x})\ ,\label{Q_0}
\eeq
where
\beq
j^{(0)}_\mu:=(\d_\nu B^\nu+m\phi)\d_\mu u- 
\d_\mu(\d_\nu B^\nu +m\phi) u\label{j^0}
\eeq
is the free BRST-current, which is conserved.
(The interpretation of the integral in (\ref{Q_0}) requires some care.) 
$Q_0$ satisfies the assumptions of the (local) 
operator BRST-formalism (Sect.~2),
in particular it is nil-potent and positive (\ref{positivity}). 
Distinguished representatives 
of the equivalence classes $[\phi]\in {\rm Ke}\,Q_0/{\rm Ra}\,Q_0$ are the 
states built up from the three spatial (two transversal for $m=0$, 
respectively) polarizations of $A$ only.
\section{Perturbative renormalization}\setcounter{equation}{0}
The starting point for a perturbative construction of an interacting 
quantum field theory is Dyson's formula for the evolution operator in the 
interaction picture. To avoid conflicts with Haag's Theorem on the 
nonexistence of the interaction picture in quantum field 
theory one multiplies the interaction Lagrangian 
${\cal L}$ with a test function $g$ and 
studies the local S-matrix
\beq
S(g{\cal L})=1+\sum_{n=1}^{\infty}\frac{i^n}{n!}
\int dx_1\ldots dx_n \, g(x_1)...g(x_n)\, T({\cal L}(x_1)\cdots {\cal L}(x_n))
\eeq
where $T$ denotes a time ordering prescription. In the limit 
$g\to 1$ (adiabatic limit) $S(g{\cal L})$ tends to the scattering matrix.
This limit, however, is plagued by infrared divergences and does not 
always exist. Interacting fields $F_{g{\cal L}}$ are obtained by Bogoliubov's formula
\beq
F_{g{\cal L}}(x)=\frac{\delta}{\delta h(x)}\vert_{h =0} S(g{\cal L})^{-1}
S(g{\cal L}+ hF) \ .\label{F_gL}
\eeq
The algebraic properties of the interacting fields within a region 
${\cal O}$ depend only on the interaction within a slightly larger region 
(Brunetti - Fredenhagen (2000)), 
hence the net of algebras in the sense of AQFT can be constructed in 
the adiabatic limit without infrared problems. (This is called the 
'algebraic adiabatic limit'.)

The construction of the interacting theory is thus reduced to a definition of 
time ordered products of fields. This is the program of causal 
perturbation theory ('CPT') which was developped by Epstein - Glaser (1973) on 
the basis of previous work by St\"uckelberg and Bogoliubov - Shirkov (1959). 
For simplicity we describe CPT for a real scalar field.
Let $\varphi$ be a classical real scalar field which is not restricted 
by any field 
equation. Let ${\cal P}$ denote the algebra of polynomials in $\varphi$ and 
all its partial derivatives $\d^a\varphi$ with multi-indices $a\in\NN_0^4$ 
and. 
The time ordered products $(T_n)_{n\in\NN}$,
are {\it linear} and {\it symmetric} maps $T_n:
(\mathcal{P}\otimes\mathcal{D}(\RR^4))^{\otimes n}\rightarrow L({\cal D})$,
where $L({\cal D})$ is the space of operators on a dense invariant domain
${\cal D}$ in the Fock space of the scalar free field. 
One often uses the informal notation
\begin{equation}
  T_n(g_1F_1\otimes...\otimes g_nF_n)=\int dx_1...dx_n\,
T_n(F_1(x_1),...,F_n(x_n))g_1(x_1)...g_n(x_n)\ ,
\end{equation}
where $F_j\in{\cal P},\>g_j\in\mathcal{D}(\RR^4)$.

The sequence $(T_n)$ is constructed by induction on $n$, starting with the 
{\it initial condition}
\beq
T_1(\prod_j\d^{a_j}\varphi(x))=\> :\prod_j\d^{a_j}\phi(x):\ , 
\eeq
where the r.h.s. is a Wick polynomial of the free field $\phi$.
In the inductive step the requirement of {\it causality} 
plays the main role, i.e. the condition that
  \beq
T_n(f_1\otimes...\otimes f_n)=
T_k(f_1\otimes...\otimes f_k)T_{n-k}(f_{k+1}\otimes...\otimes f_n)\label{caus}
\eeq
if $\bigl(\supp f_1\cup...\cup\supp f_k\bigr)\cap 
\bigl((\supp f_{k+1}\cup...\cup\supp f_n)+\bar V_{-}\bigr)=
\emptyset$ (where $\bar V_{-}$ is the closed backward light cone). 
This condition expresses the composition law for evolution operators 
in a relativistically invariant and local way.  
Causality determines $T_{n}$ as an operator valued distribution on $\RR^{4n}$ 
in terms of
the inductively known $T_l\ ,\>l<n$ outside of the total diagonal 
$\Delta_n:=\{(x_1,...,x_{n})\>|\> x_1=...=x_{n}\}$, i.e. on test functions 
from ${\cal D}(\RR^{4n}\setminus \Delta_{n})$. 

{\it Perturbative renormalization} is now the extension of $T_{n}$
to the full test function space 
${\cal D}(\RR^{4n})$. Generally, this
extension is non-unique. In contrast to other methods of 
renormalization no divergences appear, but the ambiguities 
correspond to the finite renormalizations which remain after removal of 
divergences by infinite counter terms. 
The ambiguities can be reduced by {\it
  (re-)normalization conditions}, which means that one requires that
certain properties which hold by induction on
${\cal D}(\RR^{4n}\setminus \Delta_{n})$ are maintained in the 
extension, namely:
\begin{itemize}
\item{\bf (N0)} A bound on the degree of singularity near the total diagonal. 
\item{\bf (N1)} Poincar\'e covariance.
\item{\bf (N2)} Unitarity of the local S-matrix.
\item{\bf (N3)} A relation to the time-ordered products of sub-polynomials.
\item{\bf (N4)} The field equation for the interacting field 
$\varphi_{g{\cal L}}$ (\ref{F_gL}).
\item{\bf (AWI)} The {\bf Action Ward identity} (Stora and D\"utsch - Fredenhagen (2003)):
$\d^\mu T(...F_l(x)...)=T(...\d^\mu F_l(x)...)$. 
This condition can be understood as the 
requirement that physics depends on the action only, so total derivatives 
in the interaction Lagrangian can be removed.
\item Further symmetries, in particular in gauge theories Ward identities 
expressing
BRST-invariance. A universal formulation of all symmetries which can be 
derived 
from the field equation in classical field theory is 
the {\it Master Ward Identity}
(which presupposes {\bf (N3)} and {\bf (N4)}) (Boas - D\"utsch - Fredenhagen 
(2002-2003)), see Sect.~5. 
\end{itemize}
The problem of perturbative renormalization is to construct a 
solution of all these normalization conditions. Epstein and Glaser have 
constructed the solutions of {\bf (N0)-(N3)}. 
Recently, the conditions {\bf (N4)} and  {\bf (AWI)} have been included. 
The Master Ward Identity cannot always be fulfilled, the
obstructions are the famous 'anomalies' of perturbative Quantum Field
Theory.
\section{Perturbative construction of gauge theories}\setcounter{equation}{0}
In the case of a purely {\it massive} theory the adiabatic limit 
$S=\lim_{g\to 1}S(g{\cal L})$ 
exists (Epstein - Glaser (1976)), and one may adopt a formalism due to 
Kugo and Ojima (1979)
who use the fact that in these theories the BRST charge $Q$ 
can be identified with the incoming (free) BRST charge $Q_0$ (\ref{Q_0}).
For the scattering matrix $S$ to be a well defined operator 
on the physical Hilbert space of the free theory, 
${\cal H}={{\rm Ke}\,Q_0}/{{\rm Ra}\,Q_0}$, one then has to require
\begin{equation}
\lim_{g\to 1}[Q_0,T((g{\cal L})^{\otimes n})]|_{\mathrm{ker}Q_0}=0\ .\label{[Q,S]=0}
\end{equation}
This is the motivation for introducing the condition of 
'perturbative gauge invariance' 
(D\"utsch - Hurth - Krahe - Scharf (1993-1996), see the second book
of G.~Scharf in the list below): 
According to this condition, there should exist a Lorentz  
vector ${\cal L}_1^\nu\in {\cal P}$ associated to the interaction ${\cal L}$,
such that
\beq
[Q_0,T_n({\cal L}(x_1)...{\cal L}(x_n)]=i\sum_{l=1}^n\d^{x_l}_\nu 
T_n({\cal L}(x_1)...{\cal L}_1^\nu(x_l)...{\cal L}(x_n))\ .\label{[Q,T]}
\eeq
This is a somewhat stronger condition than 
(\ref{[Q,S]=0}) but has the advantage that it can be formulated
independently of the adiabatic limit. The condition (\ref{[Q,S]=0})
(or perturbative gauge invariance) can be satisfied 
for tree diagrams (i.e.~the corresponding requirement
in classical field theory can be fulfilled). In the massive case
this is impossible without a modification of the model;
the inclusion of additional {\it physical} scalar fields 
(corresponding to Higgs fields) yields a solution. 
It is gratifying that, making a polynomial ansatz for the interaction 
${\cal L}\in {\cal P}$, perturbative gauge invariance (\ref{[Q,T]}) 
for tree diagrams, renormalizability (i.e.~the mass 
dimension of ${\cal L}$ is $\leq 4$) and
some obvious requirements (e.g.~Lorentz invariance) determine ${\cal L}$
to a far extent. In particular, the Lie algebraic structure needs not to be put in, 
it can be derived in this way (Stora (1997)). Including loop diagrams
(i.e.~quantum effects),
it has been proved that {\bf (N0)-(N2)} and perturbative gauge invariance can be 
fulfilled to all orders for massless $SU(N)$-Yang-Mills theories.

Unfortunately, in the {\it massless} case, it is unlikely that the adiabatic limit
exists and, hence, an $S$-matrix formalism is problematic. 
One should better rely on the construction of 
local observables in terms of couplings with compact support. 
But then the selection 
of the observables (\ref{A}) has to be done in terms of 
the BRST-transformation $\tilde s$
of the interacting fields. For the corresponding BRST-charge one 
makes the ansatz
\beq
\tilde Q =\int d^4x \tilde j^\mu_{g{\cal L}}(x)b_\mu(x)\ ,\quad {\cal L}=
\sum_{n\geq 1}{\cal L}_n\lambda^n\ ,\label{tilde-Q}
\eeq
where $(b_\mu)$ is a smooth version of the $\delta$-function characterizing 
a Cauchy surface\footnote{There is a 
volume divergence in this integral, which can be avoided by a spatial compactification.
This does not change the abstract algebra ${\cal F}_{\cal L}({\cal O})$.}
and $\tilde j^\mu_{g{\cal L}}$ is the interacting BRST-current (\ref{F_gL}) (where 
$\tilde j_\mu=\sum_n j_\mu^{(n)}\lambda^n$ ($j_\mu^{(n)}\in {\cal P}$)
is a formal power series with $j_\mu^{(0)}$ given by (\ref{j^0})). A crucial 
requirement is that $\tilde j^\mu_{g{\cal L}}$ is conserved
in a suitable sense. This condition is essentially equivalent to perturbative 
gauge invariance and 
hence its application to classical field theory determines 
the interaction ${\cal L}$ in the same way, and in addition the deformation
$j^{(0)}\rightarrow\tilde j_{g{\cal L}}$. The latter gives also 
the interacting BRST charge and transformation,  $\tilde Q$ and $\tilde s $, 
by (\ref{tilde-Q}) and (\ref{[Q,F]}). Mostly the so obtained 
$\tilde Q$ is nil-potent in classical field theory (and hence this 
holds also for $\tilde s$).
However, in QFT conservation of $\tilde j_{g{\cal L}}$ and 
$\tilde Q^2=0$ require the validity of additional Ward identities,
beyond the condition of perturbative gauge invariance (\ref{[Q,T]}).
All the necessary identities can be derived from the 
Master Ward Identity
\beq
T_{n+1}(A,F_1,...,F_n)=-\sum_{k=1}^n T_n(F_1,...,\delta_A F_k,...,F_n)\ ,
\eeq
where $A=\delta_AS_0$ with a derivation $\delta_A$.
The Master Ward Identity is closely related to the Quantum Action Principle
which was formulated in the formalism of generating functionals of 
Green's functions. In the latter framework the anomalies have been 
classified by cohomological methods. The vanishing of anomalies of 
the BRST symmetry is a selection criterion for physically acceptable models.

In the particular case of QED, the Ward identity
\beq
\d_\mu^y T\Bigl( j^\mu (y)F_1(x_1)...F_n(x_n)\Bigr)=
i\sum_{j=1}^n\delta (y-x_j)
T\Bigl( F_1(x_1)...(\theta F_j)(x_j)...F_n(x_n)\Bigr)\label{QED}
\eeq
for the Dirac current $j^\mu :=\bar\psi\gamma^\mu \psi$,
is sufficient for the construction, where 
$(\theta F):= i(r-s)F$ for $ F=\psi^{r}\bar\psi^{s}B_1...B_l$
($B_1,...,B_l$ are non-spinorial fields) and 
$F_1,...,F_n$ run through all sub-polynomials of ${\cal L}=j^\mu A_\mu$, 
{\bf (N0)-(N4)} and (\ref{QED}) can be fulfilled to all orders (D\"utsch - Fredenhagen (1999)).

\medskip
\begin{center}
{\Large\bf Further Reading}
\end{center}
\begin{itemize}
\item Becchi, C., Rouet, A., and Stora, R., 
{\it Commun. Math. Phys.} {\bf 42}, 127 (1975);
{\it Annals of Physics (N.Y.)} {\bf 98}, 287 (1976)

\item Bogoliubov, N.N., and Shirkov, D.V., {\it "Introduction to the 
Theory of Quantized Fields"}, New York (1959)

\item D\"utsch, M., and Fredenhagen, K.,
"A local (perturbative) construction of observables 
in gauge theories: the example of QED", 
{\it Commun. Math. Phys.} {\bf 203}, 71 (1999)

\item D\"utsch, M., and Fredenhagen, K., ``The Master Ward Identity 
and Generalized Schwinger-Dyson Equation in Classical Field Theory'', 
{\it Commun. Math. Phys.} {\bf 243} 275 (2003)

\item D\"utsch, M., and Fredenhagen, K., ''Causal perturbation
  theory in terms of retarded products, and a proof of the Action 
Ward Identity'', hep-th/0403213

\item Epstein, H., and Glaser, V.,
{\it Ann. Inst. H. Poincar\'e A} {\bf 19}, 211 (1973)

\item Henneaux, M., and Teitelboim, C., {\it ``Quantization
of Gauge Systems''}, Princeton University Press (1992)

\item Kugo, T., and Ojima, I., 
{\it Suppl. Progr. Theor. Phys.} {\bf 66}, 1 (1979)

\item O. Piguet, S. Sorella, ``Algebraic renormalization: 
Perturbative renormalization, symmetries and anomalies'', Berlin: 
Springer, Lecture notes in physics (1995) 

\item Scharf, G.,
{\it "Finite Quantum Electrodynamics. The causal approach"}, 
2nd. ed., Springer-Verlag (1995)

\item Scharf, G., {\it ``Quantum Gauge Theories - 
A True Ghost Story''}, John Wiley and Sons (2001)

\item Weinberg, S., {\it ``The Quantum Theory of Fields''}, volume II,
Cambridge University Press (1996)
\end{itemize}

\end{document}